\documentstyle[12pt]{article}
\def\be{\begin{equation}}
\def\ee{\end{equation}}
\def\bea{\begin{eqnarray}}
\def\eea{\end{eqnarray}}

\begin{document}
\pagestyle{empty}
\begin{flushright}
{BROWN-HET-1019}\\
{PAR-LPTHE 95-48}\\
{December 1995}
\end{flushright}
\vskip .15in
\begin{center}
 COLLECTIVE FIELD THEORY OF THE MATRIX-VECTOR MODELS\\
\vskip .15in
by\\
\vskip .15in
Jean AVAN\footnote{Permanent address:  LPTHE Paris 6, CNRS URA 280;
Box 126, 4 Place Jussieu, F-75252, Paris Cedex 05.}\\
\vskip .05in
and\\
\vskip .05in
Antal JEVICKI\\
{\it Department of Physics}\\
{\it Brown University}\\
{\it Providence, Rhode Island  02912  USA}\\
\vskip .25in
{\bf ABSTRACT}
\end{center}
We construct collective field theories associated with one-matrix plus
$r$ vector models. Such field theories describe the continuum
limit of spin Calogero Moser models. The invariant collective fields consist 
of a scalar
density coupled to a set of fields in the adjoint representation of $U(r)$.
Hermiticity conditions for the general quadratic Hamiltonians
lead to a new type of extended non-linear algebra of differential operators
acting on the Jacobian. It includes both Virasoro and  $SU(r)$ 
(included in $sl(r, {\bf C})
\times sl(r, {\bf C})$) current algebras. A systematic construction of exact
eigenstates for the coupled field theory is given and exemplified.

\vskip .10in
\newpage
\setcounter{page}{1}
\pagestyle{plain}

\section{Introduction}
\medskip
Numerous studies have been devoted to the problem of collective field theory of
matrix models.  The one-matrix Hamiltonian problem is by now well understood
in this approach \cite{JS}. The underlying $w_{\infty}$ algebra of observables
plays the role of a spectrum-generating symmetry \cite{AJ1}. The connection
to quantum Calogero Moser models was also extensively studied
\cite{AJL,MP1,Iso,Awa}.   Recently
 various generalizations, related to multimatrix models \cite {JR,Ike,SY,Za}
have  been considered. The relevant space of
observables was shown to be the space of marked loops acted upon by splitting
and joining operators. The difficulty of manipulating non-local operators
on such a huge space and the complexity of the relevant algebraic structures
has lead to introducing simplifications of this approach \cite{JR,Ike}.

We investigate here a set of theories intermediate between the one- and
multi-matrix models.  We are going
to construct the collective field theory for a dynamical system
defined on a phase
space comprising a hermitian matrix variable $ m_{ij}, i,j = 1\cdots N $
or equivalently its
unitary exponential $M \equiv {\rm exp}\,\, im$ with a conjugate
momentum field $p_{ij}$, plus a set of $r$ complex vector variables
$\{ x_i^a , a = 1 \cdots r; i = 1\cdots N\}$  and their conjugate
momenta $y_i^a = {\partial\over \partial x_i^a}$. The group $SU(N)$ has a
natural hamiltonian action on this phase space, and we may therefore
define collective models with a Hilbert space consisting
of functions of the $SU(N)$-invariant variables
$\{ Tr M^n \equiv \phi_n^0 ,
Tr (\bar{x}^a M^n  x^b) \equiv \psi_n^{ab}, a,b = 1
\cdots r\}$.
The variables $\psi^{ab}$ now carry two $U(r)$ indices, and it will be seen
that the field theories
which we develop here will describe a consistent interaction of
a collective boson $\phi$ with  $U(r)$ current algebra degrees 
of freedom. 
These models have a number
of interesting applications.  First of all and most obviously,
they are toy models for interacting quarks and open strings
\cite{Aff,CIT,MP2,KK,Mo}.
Then, as we will comment, suitable
restrictions of the Hilbert space of wave functions for these collective
field theories could ultimately provide us with a consistent reduction of
any matrix model.

 Finally, the most relevant feature of this model, from our present point of
view, is that, by Hamiltonian reduction the
free matrix Laplacian
$ H = \sum \frac{\partial ^2}{\partial m_{ij} \partial m_{ji}}$ acting
on the matrix-vector configuration space induces the well-known spin (or Euler)
Calogero-Moser models \cite{GH,Ne} and the related 
Haldane-Shastry models \cite {HH,HT}. Many results were recently obtained
on these various models, in particular
their exact Yangian symmetry \cite{BGHM,BAB,Sc} and the construction of
commuting quantum Hamiltonians by the $R$ matrix method
\cite{ABB} for the Euler Calogero Moser case.

This connection to spin Calogero-Moser models needs to be made more precise.
The phase space here contains $r$ complex vector fields and their canonically
conjugate (but not hermitean conjugate) momenta. The action of the group
$SU(N)$ on the vector fields leads to a contribution
to the moment map taking the form, after quantization,
of a particular representation of $SU(N)$:

\be
F_{ij} = \sum_{a=1}^{r} \, x_a^i  \, {\partial\over\partial x_{a}^{j}} -
\sum_{b=1}^r \, \bar{x}_b^j  \, {\partial\over \partial\bar{x}_b^j}
\label{3}
\ee

These operators, whichever way they are represented, always
define the ``spin'' interaction of the quantum Euler Calogero Moser models
$F_{ij}F_{ji}$ \cite{ABB}.
In this representation the diagonal generators $F_{ii}$ annihilate all
collective variables
$\phi_n^0 , \phi_m^{ab}$, thereby realizing on the reduced $SU(N)$-invariant
Hilbert space of wave functions the ``zero-weight condition''
necessary for integrability of any quantum spin Calogero-Moser model
\cite{ABB}.
The fact that it is based on two conjugate vector representations of $SU(N)$
indicates however a closer connection with the integrable
$N\otimes \bar{N}$
Calogero-Moser quantum system constructed in \cite{ABB}.

The standard spin Calogero Moser Hamiltonian \cite{GH} follows instead from the
Hamiltonian reduction of a free theory on a phase space containing one set
of $r$ complex vectors together with their canonical {\it and} hermitian
conjugate momenta. These vectors ought therefore to be parametrized as
$a^b_k = x^b_k + ip^b_k ; a_k^{b \dagger} = x^b_k -i p^b_k$ and the group
$SU(N)$ acts on the vectors $a^b_k ,a_k^{b \dagger}$, with a contribution to
the moment map taking the form of another, oscillator-like,
representation of $SU(N)$:

\be
F_{ij} = \sum_{b=1}^{r} a_i^{b \dagger} a^b_j
\nonumber
\ee

The resulting field-theoretic algebraic structures of these models are however
identical.

Based upon these introductory remarks our investigations run as follows.
In a first part we determine the set of equations obeyed by the Jacobian
of the change of variables from the original configuration space
$\{ M, x_a^i\}$ to the adjoint-invariant configuration space
$\{ \phi_n^0 , \phi_m^{ab}\}$.
The differential operators acting on $J$ close a non-linear algebra,
which we shall obtain explicitly. From the structure of the discrete
Hamiltonian we also derive the expected form of the
non-hermitian continuum Hamiltonian 

In a second part we examine the spectrum of the Hamiltonian corresponding
 to the reduction of the free matrix Laplacian.
We indicate how the structure of the reduced Hamiltonian allows a
systematic and exhaustive computation of its eigenvectors and eigenvalues
inside the Hilbert space of polynomials in $\{ \phi_n^0 , \phi_m^{ab}\}$,
and describe fully the diagonalizing procedure. We explicitly apply
this procedure to obtain some of the simplest eigenstate with low-lying
eigenvalues.  We justify the form of both eigenstates and eigenvalues
from purely group-theoretical arguments connecting the representation theories
of $SU(N)$ and of permutation groups $S_p$.

 Finally we give a derivation of the hermitian
collective field theory at the classical
level for the spin Calogero Moser model. The interaction
between the scalar and spin-type fields is brought in by the Poisson
structure.
This derivation also provides us with a semi-classical
limit of  matrix-vector collective theories, without the technical
complications due to the ordering problems and the conjugation by the Jacobian
required to get the exact quantum hermitian Hamiltonian. The basic algebraic
structure of the Hamiltonian is thus made more transparent.

\medskip
\section{The Collective Field Theory for the Matrix-Vector Model}
\medskip

We apply to our configuration space the general formalism defined in \cite{JS}.
 The central issue is the rewriting of an original hermitian Hamiltonian
written in terms of  variables $(M)_{ij} , \bar{x}_i , x_j$ as an operator in
terms of
collective variables
\be
\phi_n =  Tr (M^n ) \;\; ; \;\;  \psi_n^{ab}  =  \bar{x}^a \cdot M^n \cdot
 x^b \nonumber
\ee

It is appropriate to indicate here in what sense this series of
models may be seen as consistent truncations of a two-matrix problem. 
 It is known
(see for instance \cite {Za}) that the full two-matrix problem leads to an
intricate type of observable space known as
loop space with loops parametrized by words
of arbitrary finite length:
\be
Tr \left( U_1^{n_{1}} U_2^{n_{2}} U_1^{n_{1}^{'}} U_2^{n_{2}^{'}}
U_1^{n_{1}^{''}} \cdots \right)
\ee
This space can be reduced to polynomials in the variables
$\phi_n = Tr (U_1^n )$ and $\psi_n^{ab} = Tr ( \bar{x}_b x_a U_1)$ by
representing $U_2$ as $ (U_2)^{ij} \equiv  \sum_{a=1}^{r} \bar{x}_a^{i} x_a^j +
\delta ^{ij}$ for some finite $r$.
Not every polynomial in this reduced set of
variables follows from the reduction of observables of the original
multi-matrixmodel.  There is a natural action of $U(r)$ on the variables
$x,\bar{x}$
viewed
as $N$ $r$-component vectors, and $U_2$ is invariant under $U(r)$. It
easily follows that 
the polynomials $P (\phi_n^0 , \phi_m^{ab} )$ invariant under the adjoint
action of $U(r_2) \times \cdots U(r_p )$ on $\psi_m^{ab}$ are the observables
of the original multimatrix model.

  To derive the collective field theory for the matrix-vector case, we follow 
the familiar procedure \cite{JS}. This essentially requires, beyond an
obvious chain rule for derivatives, the introduction of the Jacobian $J$
associated with the change of variables, in order to define the new measure
with respect to which the collective Hamiltonian is hermitian.  Differential
equations obeyed by $J$ then follow and it is in this context that a closed
operator algebra arises as Frobenius-Schur conditions on the compatibility
of the differential operators acting upon $J$.

Starting from the Hamiltonian, assumed to be at most quadratic in the
derivatives ${\partial\over\partial m_{ij}}$ and ${\partial\over
\partial x_i^a} $,
${\partial\over \partial \bar{x}_i^b}$, with the most general $U(N)$ invariant
form
\bea
 H & = &\sum_{i,j}^N {1\over 2} \, m_M \, {\partial^2\over \partial m_{ij}
\partial m_{ji}} + \sum_{a=1}^r \sum_{i=1}^N \, {1\over 2} \, m_a \,
{\partial\over \partial\bar{x}_a^i} \, {\partial\over \partial x_a^i}
\nonumber \\
&+ & {\rm potential \,\, terms} \label{6}
\eea
one applies it to a wave function $\Psi (\phi_n , \psi_m^{ab} )$. 

For the Hamiltonian given by the matrix Laplacian we obtain (setting $m_M =1$):
\bea
H_1 &=& \frac{1}{2} \sum_{n,m} nm \, \phi_{n+m} \, {d^2\over d\phi_n d\phi_m} +
\sum_{n,m,ab} \, nm \,\psi_{n+m}^{ab} \, {d^2\over d\phi_n d\psi_m^{ab}}
\nonumber\\
&+& \frac{1}{2} \sum_{n,m; ab,cd} \, \sum_{u=0}^{n+m} F (n,m,u) 
\psi_{n+m-u}^{ad}
 \psi_u^{cb} \, {d^2\over d\psi_n^{ab} d\psi_m^{cd} }\nonumber\\
&+ & \frac{1}{2}\sum_n \, n \sum_{u=0}^n \, \phi_{n-u} \phi_u \, {d\over d\phi_n} +
 \sum_n \sum_{u=0}^n \, (n-u) \phi_u \psi_{n-u}^{ab} \, {d\over d\psi_n^{ab}}
\label{8}
\eea
with
\bea
F (n,m,u) & = & u \, \, {\rm for}\,\, 0\leq u \leq \inf  (n,m)\nonumber\\
F(n,m,u) & = & \inf (n,m) \,\,{\rm for} \,\, \inf (n,m) \leq u \leq
\sup (n,m)\nonumber\\
F (n,m,u) & = & n+m-u \,\,{\rm for} \,\, \sup (n,m) \leq u \leq n+m
\nonumber
\eea
while for the vector Laplacian we get:
\be
H_2 = \sum_{a=1}^r \, {m_a\over 2} \left\{  \sum_{m,n,cd} \,
\psi_{m+n}^{cd}
 \, {d^2\over d\psi_m^{ca} d\psi_n^{ad} }
 + \sum_m \phi_m \, {d\over d\psi_m^{aa}}
 \right\}
\label{9}
\ee

In (\ref{8}) $F(n,m,u)$ is in fact a summation kernel and the corresponding
term in (\ref{8}) can be rewritten as

\bea
&&\sum_{u=0}^{n+m} \, F(n,m,u)\cdot =\sum_{u=0}^{\inf(n,m)} \, u \cdot +
\sum_{\sup (n,m)}^{\sup (n,m)} \, \inf (n,m) \cdot \nonumber\\
&& + \sum_{\sup (n,m)}^{n+m} (n+m-u)\, .
\eea

\noindent where the sums are understood algebraically, i.e. $\sum_a^b = -
\sum_b^a$ if $ b\leq  a$.

Let us now recall the basic proposition of collective field theory \cite{JS}:

Given a Hamiltonian $H$ of the form:
\be
H = \sum_k \left\{ \sum_{k'} \, \Omega_{kk'} \, {d\over d\phi_{k'}} + \omega_k
\right\} {d\over d\phi_k} + V (\phi_k )
\label{12}
\ee

\noindent the Jacobian which makes $H$ hermitian by conjugation is to obey a
set of equations:
\be
\sum_{k'} \left\{ \Omega_{kk'} {d\over d\phi_{k'}} + \omega_k\right\}^{\dag} J
= 0 \label{13}
\ee

\noindent where ($\dag$) denotes the hermitian conjugation for differential
 operators.  The hermitian Hamiltonian then takes the form

\bea
H^h &=& \sum_{k, k'} \, {d\over d\phi_k} \, \Omega_{kk'} \, {d\over d\phi_{k'}}
- {1\over 2} \, {d^2 \Omega_{kk'}\over d\phi_k d\phi_{k'}} +
 V (\phi_k ) -
{1\over 4} \, \sum_k \, {\partial \omega_k\over \partial\phi_k }\nonumber\\
&+& {1\over 4} \sum_{i,j} \left\{ \omega_i + \sum_k \,
{\partial \Omega_{ik}\over \partial\phi_k} \right\}
\left( \Omega^{-1} \right)_{ij} \left\{ \omega_j +
\sum_k \, {\partial\Omega_{ik}\over \partial\phi_k}\right\}
\label{14}
\eea
\newline

In order for this scheme to be applicable, the set of differential
 operators (\ref{13}) must close an algebra or at least generate a closed
algebra of constraints under successive commutations.  Examples of this
construction are the one-matrix case realizing a Virasoro algebra \cite{JS,JR},
the general loop space algebra \cite{SY,Za} and the Kac-Moody algebra
 considered
in \cite{AJ2}. In this last case the corresponding collective theories were
constructed by postulating a Kac-Moody type algebra of constraints and
inverting the canonical procedure, thereby going from (\ref{13}) to (\ref{12})
and (\ref{14}).

In order to get the algebra of differential operators annihilating the
Jacobian, we write $H$ as:
\bea
H & = & {1\over 2} m_M \sum_n n O_n^0 \, {d\over d\phi_n} + {1\over 2} m_M
\sum_{n,ab} n\left\{ \tilde{O}_n^{ab} - \sum_{u=0}^{n-1} \left( 1 - {u\over n}
\right)
O_{n,u}^{ab} \right\} \, {d\over d\psi_n^{ab}}\nonumber\\
&+& {1\over 2} \sum_{n,a,b} \left\{ m_a \, J_n^{ab} + m_b \bar{J}_n^{ab}
\right\} {d\over d\psi_n^{ab}}
\label{15}
\eea
where one defines:
\be
O_n^0 = \sum_q  q\phi_{q+n} {d\over d\phi_q} + \sum_{q,cd} \, q \,
\psi_{q+n}^{cd} \, {d\over d\psi_q^{cd}} + \sum_{q=0}^n \, \phi_{n-q} \phi_q
\ee
\be
\tilde{O}_n^{ab} = \sum_q q \psi_{q+n}^{ab} \, {d\over d\phi_q} + \sum_{q,cd}
\sum_{u=0}^{q-1} \, \psi_{n+q-u}^{ad} \, \psi_u^{cb} \, {d\over d\psi_q^{cd}} +
\sum_{u=0}^{n-1} \, \phi_{n-u} \psi_u^{ab}
\ee
\be
J_n^{ab} = \sum_{m,c} \psi_{m+n}^{cb} \, {d\over d\psi_m^{ca}} + \delta^{ab}
 \phi_m \quad ; \quad \bar{J}_n^{ab} = \sum_{m,c} \psi_{n+m}^{ac} \, {d\over
d\psi_n^{bc}} + \delta^{ab} \phi_m
\ee
\be
O_{n,u}^{ab} = - \sum_d \psi_{n-u}^{ad} \, J_u^{bd} + \sum_c \psi_u^{cb} \,
\bar{J}_{n-u}^{ac}
\ee

One introduces the combined operator:
\be
O_{n}^{ab} \equiv \tilde{O}_{n}^{ab} + \sum_{u=0}^{n-1} \psi^{ad}_{n-u}
 J^{db}_{u}
\label{extra}
\ee
in order to have a more symmetric cubic term $\sum_{q,cd}
\sum_{u=0}^{n+q-1} \, \psi_{n+q-u}^{ad} \, \psi_u^{cb} \, {d\over d\psi_q^{cd}}$

The Jacobian $J$ is independent of which particular Hamiltonian is chosen;
 hence it should not depend on $m_M$ and $m_a$.  It follows that it must be
annihilated separately by $O_n^0 ; O_n^{ab} ; J_n^{ab}; \bar{J}_m^{ab}$ and
 $O_{n,u}^{ab}$. The last is a dependent operator. Notice also that conjugation
properties of $\psi$ and $M$ imply that $J^{ab}_n = \bar{J}^{ba}_n$.

\noindent The conjugate operators to $\{ O_n^0 , O_n^{ab}, J_n^{ab} ,
\bar{J}_n^{ab}\}$ do close a non-linear algebra with a number of
characteristic features:
\bea
\left[ O_n^0 , O_m^0 \right] & = & (m-n) \, O_{n+m}^0 \quad\quad
 ( {\rm Virasoro\,\, algebra\,\, of\,\, 1-matrix\,\, model})\nonumber\\
\left[ O_n^0 , \bar{J}_m^{ab} \right] & = & m \bar{J}_{m+n}^{ab}\nonumber\\
 \left[ O_n^0 , J_m^{ab} \right] & =  &m J_{m+n}^{ab}
\qquad\qquad\quad\qquad\quad ({\rm Spin\,\, 1\,\, currents})\nonumber\\
&&\nonumber\\
\left[ J_n^{ab} , J_m^{cd} \right] & = & -\delta^{bc} \, J_{n+m}^{ad} +
\delta^{ad} \, J_{n+m}^{cb} \nonumber\\
\left[ \bar{J}_n^{ab} , \bar{J}_m^{cd} \right] & = & \delta^{bc}
\bar{J}_{n+m}^{ad} - \delta^{ad} \, \bar{J}_{n+m}^{cb} \qquad (sl(r, {\bf C})
 \times sl(r, {\bf C})
\,\, {\rm current\,\, algebra} )\nonumber\\
\left[ J_n^{ab} , \bar{J}_m^{cd} \right]& = &0 \nonumber\\
&&\nonumber\\
\left[ O_n^0 , O_m^{ab} \right] & = & (m-n) \, O_{m+n}^{ab} + \sum_{u=0}^{n-1} 
(n-u) \left( \sum_{g} \psi^{gb}_{u} \bar{J}_{n+m-u}^{ag} + \sum_{h}
\psi^{ah}_{u} J^{hb}_{n+m-u} \right)  \nonumber\\
\left[J_n^{ab} , O_m^{cd} \right] & = & \delta^{ad} \, O_{n+m}^{cb}\nonumber\\
&&  - \sum_{u=0}^{n-1} \left\{ \psi_{u}^{cb} \, J_{n+m-u}^{ad} + \delta^{ad}
 \sum_f  \psi_u^{fb} \, \bar{J}_{m+n-u}^{cf} \right\} \nonumber\\
\left[ \bar{J}_n^{ab} , O_m^{cd}\right] & = & \delta^{bc} \, O_{n+m}^{ad}  -
\sum_{u=0}^{n-1} \, \left\{ \psi_u^{ad} \, \bar{J}_{n+m-u}^{cb}
+\delta^{bc} \sum_f \psi^{af}_{u} J^{fd}_{n+m-u} \right\}
\nonumber\\
\left[ O_n^{ab} , O_m^{cd} \right]& = & \sum^{m-1}_{v=0} \sum_{u=0}^{m-v-1}
\psi^{cb}_{u} \sum_{f} \psi^{fd}_{n+m-u-v} \bar{J}^{af}_{v} + \sum^{n-1}_{v=0} 
\sum_{u=0}^{n-v-1} \psi^{cb}_{v} \sum_{h} \psi^{ah}_{n-u-v} J^{hd}_{n+u} \;\;\;\;\;\;\;\;\;
\nonumber\\
 &-&\sum^{n-1}_{v=0} \sum_{u=0}^{n-v-1}
\psi^{ad}_{u} \sum_{f} \psi^{fd}_{n+m-u-v} \bar{J}^{cf}_{v}
- \sum^{m-1}_{v=0} \sum_{u=0}^{m-v-1}
\psi^{ad}_{v} \sum_{h} \psi^{ch}_{m-u-v} J^{hb}_{n+v}
\eea

\noindent

The closure of this constraint algebra now ensures the existence of the
Jacobian from Frobenius-Schur theorem, at least locally. 

This algebra realizes a non-Abelian extension of the simplest case $r=1$
studied in \cite{Mo}.  This structure is of an original
type. It contains a linear current algebra $sl(r, {\bf C}) \times
sl(r, {\bf C})$ and a Virasoro algebra. Note that this chiral current
algebra contains a current algebra $SU(r)$ generated by ${\cal J}^{ab}_n
\equiv J^{ab}_n - \bar{J}^{ab}_n$. This current algebra  plays a crucial
role in the continuum field theory.
The non-linear commutators containing the generators $O^{ab}_n$ however
do not take the form of $W$-algebra commutators; instead of closing over
the product of the Virasoro generator with itself they close on the product
of bilinear of the fields with the two $sl(r, {\bf C})$  currents. 
  
 The hermitian form for
the Hamiltonian is obtained from eq. (\ref{14}); however the matricial nature
of the kernel $\Omega_{ij}$ makes it difficult to give an explicit global
formula for any value of $r$.  Of course $H$ can be computed systematically for
each value $r$.

Let us summarize the basic features of the field theory which we have 
constructed and compare it with previous works.
The basic Hamiltonian $H_1$ representing the matrix Laplacian, and after
reduction the spin Calogero Moser model, consists of three terms:

{\bf 1)} The  scalar field self-interaction:

\be
 h_1 = \sum_{n,m} \phi_{n+m} \frac{d^2}{d\phi_n d\phi_m} +
 \sum_n n \sum_{u=0}^n \phi_{n-u} \phi_u \frac{d}{d \phi_n}
\ee
representing the well-known one-matrix collective Hamiltonian \cite{AJ1,JR}.

{\bf 2)} The pure  $SU(r)$ current algebra term:
\be
 h_2 = \sum_{n,m} \sum_{ab,cd} \sum_{u=0}^{n+m} F(n,m,u) \psi^{ad}_{n+m-u}
\psi^{cb}_u \frac{d^2}{d \psi^{ab}_n \psi^{cd}_m}
\ee

which has the same overall form as the $SU(r)$ 
current-current interaction proposed in 
\cite{HT} as a conformal field version of the pure spin Haldane-Shastry
model. 

{\bf 3)} The coupling between the bosonic scalar field $\phi$ and the 
current-algebra:

\be
 h_3 = \sum_{n,m,ab} nm \psi^{ab}_{n+m} \frac{d^2}{d \phi_n d \psi^{ab}_m}
+ \sum_n \sum_{u=0}^{n} (n-u) \phi_u \psi^{ab}_{n-u} \frac{d}{d \psi^{ab}_n}
\ee

The current-algebra operator coupled to the scalar field is recognized
as the energy-momentum tensor $T^J (z)$ of the current ${\cal J}^{ab}$.
 Interestingly
the same scalar-to-Virasoro-generator coupling occurred in \cite{AJ2}, in the
context of constructing a field theory of the one-matrix model incorporating
the topological ``discrete''states.

The discrete non-hermitian Hamiltonian (6) can be consistently restricted
to the positive-index modes for $\phi$ and $\psi$. Introducing now
the conformal field notation :

\bea
\Phi(z) = \sum_{n \geq 0} z^{n-1} \phi_n &;&
\Pi(z) = \sum_{n \geq 0} z^{-n} \frac{d}{d \phi_n} \nonumber\\
\alpha (z) = \Phi(z) + \partial_z \Pi(z) &;&
{\cal J}^{ab} = \sum_{n \in {\bf Z}} {\cal J}^{ab}_n z^{-n-1} 
\eea

the resulting Hamiltonian reads:

\be 
H_1 = \int dz \; z^2 \left( \frac{( \alpha (z))^3}{6} +  \alpha (z)
T^J (z) \right) + \int dz \int dy \sum_{ab} 
\frac{{\cal J}^{ab} (z) {\cal J}^{ba} (y)}{(z-y)^2}
\label{ham2}
\ee

it represents a continuum field theory of a collective boson $\alpha$
interacting with a $SU(r)$ 
current algebra. It is known from \cite{Sc,AN} that a more
careful definition of the pure current algebra term requires introduction of
a cubic term in the continuum theory from considerations of symmetry. 
These terms could arise, in our present approach, from effects of 
the non-trivial Jacobian.

\medskip
\section{The Diagonalization of $H_1$}
\medskip

The Hamiltonian
\be
H_1 = \sum_{i,j}\, {\rm Tr}\, (e_{ij} U {\partial\over\partial U} ) \cdot
 {\rm Tr}\, (e_{ji} U {\partial\over\partial U} ) = \sum_{i,j}
{\partial^2\over\partial m_{ij} \partial m_{ji}}
\ee
is  of particular interest since it reduces under the well-known 
Marsden-Weinstein procedure
 to the spin Calogero-Moser system. We shall study the spectrum of its
restriction to $U(N)$-invariant quantities of the configuration space
$\{ M, x_i^a , \bar{x}_j^b\}$. Since such quantities are obtained by
elimination of the angular degrees of freedom of the configuration space,
this spectrum should be included (not identical, due to the extra 
permutation symmetry of
involved in defining the $\phi$-variables) into the spectrum of a spin 
Calogero-Moser system with
complex vector variables. Note that the Hamiltonian derived in \cite{ABB} from
an $N\times\bar{N}$ tensor representation of the Yang-Baxter equation may be
interpreted as  a restriction to a smaller Hilbert space of this type of
Calogero-Moser
model.

 We will use the nonhermitean collective discrete Hamiltonian 
(\ref{8}) derived in the
previous section. Its spectrum is the same as the
spectrum of its hermitian version, and the eigenstates are obtained a
multiplication by the square root of the Jacobian.

\noindent The strategy for diagonalizing $H_1$ in (\ref{8}) proceeds from three
properties of $H_1$:
\begin{description}

\item{{\bf Property 1:}}  The number of vector variables $\psi_n^{ab}$ is
conserved.  This is due to the global U(1) symmetry of (\ref{8}) $\psi_n^{ab}
\rightarrow \lambda \psi_n^{ab}$, itself stemming from the trivial chiral
symmetry of $H_1$  (which does not act on $x, \bar{x}$) under $x_a \rightarrow
\lambda x_a , \bar{x}_b \rightarrow \bar{\lambda} \bar{x}_b$.  It follows that
 eigenfunctions are separated into selection sectors with a given number $N_V$
 of vector variables.

\item{{\bf Property 2:}}  the total sum of indices $n_i$ in $\phi_{n_{i}} ,
\psi_{n_j}^{ab}$ is conserved.  This follows from the form of $H_1$ as
 ${\rm Tr}\, (U \, {\partial\over \partial U} \cdot U
{\partial\over \partial U} )$ which conserves the total number of $U$-type
 variables.  This provides us with a further separation of eigenfunctions
into selection sectors with a given number $N_0$ of $U$-variables.

\item{{\bf Property 3:}}  Define $P ( N_0^+ , N_0^- ) = \{ {\rm Polynomials}
 ( \phi_n^0 , \phi_m^{ab}) , \sum_{n,m>0} n_i + m_i = N_0^+ ,\,\,
 \sum_{n,m<0} n_j + m_j = - N_0^- \}$.  The flag vector spaces defined as
 $F(N_0 , q_0 ) = \bigoplus_{q=0}^{q_{0}\geq 0} \, P (N_0 + q , q )$
 for
$N_0 \geq 0 $ and $ \bigoplus_{q=0}^{q_{0}\geq 0 }\,
P(q , - N_0 + q ) $
 for $N_0 \leq 0$, are invariant under $H_1$.  In particular the
polynomials in purely positive or negative index variables $\phi, \psi^{ab}$,
 are invariant under $H_1$.
\end{description}

The proof of Property 3 is obtained by recursion over $q_0$.  The choice
$N_0 \geq 0$ or $N_0 \leq 0$ is arbitrary, in fact the eigenvalues and
eigenvectors in sectors $N_0 \leq 0$ are deduced from sectors $N_0 \geq 0$ by
changing $U$ into $U^{-1}$; this keeps the form of $H_1$ and changes the
indices $n_i$; hence eigenfunctions have same eigenvalues with signs of indices
$n_i$ globally changed.

\noindent  We choose $N_0\geq 0$.  When $q_0= 0$, it is immediate by inspection
of (\ref{8}) that $H_1$ generates only positive indices out of positive
indices, hence $P ( N_0, 0)$ is stable.  Assume Property 3 to be proved up to
$q_0$.
 Consider $P (N_0 + q_0 + 1 , q_0 + 1 )$.  From (\ref{8}), we see that

\begin{description}

\item{a)} terms linear in derivative pick one index and split it into
same-sign pair of indices, hence both $N_0^+$ and $N_0^-$ are conserved.

\item{b)}  the first two quadratic terms pick pairs $(n,m)$ and join them as
$n+m$.  If $n,m$ have same sign, this conserves both $N_0^+$ and $N_0^-$; if
not, $N_0^+$ and $N_0^-$ are simultaneously decreased.  Hence one ends up in
 $P (N_0 + q' , q'$) with $0\leq q' < q_0 + 1$, which is inside the flag
vector space.

\item{(c)}  the third quadratic term picks pairs $(n,m)$ and turns them into
pairs $(n', m')$.  If $n,m$ have same sign, $n'$ and $m'$ keep that same sign
 and same sum, hence both $N_0^+$ and $N_0^-$ are conserved. If not,
say $n \geq 0 \geq m$, $n'$ and $m'$ stay between $n$ and $m$ with same sum.
 Hence the positive term (if any) is smaller than $m$; the negative term, if
 any, is smaller in absolute value than $\vert m \vert$; and the sum (if both
$n', m'$ have same sign) is $n+m$ which is necessarily smaller, in absolute
value, than the single index $n$ or $m$ which it replaces in $N_0^+$ or
$N_0^-$.  Therefore the third quadratic term sends $P (N_0 + q_0, q_0 + 1 )$
either to itself, or to a lower-$q_0$ vector space, stabilizing the flag
 vector space $F(N_0 , q_0)$.

\end{description}

\noindent The diagonalizing procedure now follows:
\begin{enumerate}
\item  Fix the values of $N_0$ (taken to be positive) and $N_V$.

\item Diagonalize $H_1$ on the finite vector space $P (N_0, 0)$.  This shall
 be described in more detail later.

\item Diagonalize $H_1$ by recursion on the flag spaces $F (N_0,q_0)
\supset \, F (N_0,q_0-1)\supset\cdots \supset F (N_0,0) = P (N_0,0)$.
 The recursion is made possible by the nested structure of $H_1$ exhibited in
the proof of Prop. 3, namely $H_1 (P (N_0+q_0,q_0))\subset
P (N_0+q_0,q_0) \bigcup_{q'<q_{0}} \, P (N_0+q,q')$.

\end{enumerate}

\noindent Moreover it follows that one can obtain the spectrum by restricting
the matrix of $H_1$ to the sole matrix elements inside the space
$P (N_0 + q_0 , q_0)$.  Indeed, diagonalizing this smaller
$P (N_0 + q_0 , q_0 )\rightarrow P (N_0 + q_0 , q_0 )$ matrix leaves the full
 $F (N_0 , q_0 ) \rightarrow F(N_0 + q_0 ) $
      matrix as:
\bea
H_1 \vert V_n \in P (N_0 + q_0 , q_0 ) > & = & \epsilon_n \vert V_n > + \sum
\lambda_{n'} \vert V_n' >\in P (N_0 + q' , q' ), q' <g_0  \nonumber\\
H_1 \vert V_{n'} >& = & \epsilon_{n'} \vert V_{n'} > \quad {\rm (recursion
\,\, hypothesis)}
\eea
from which the exact eigenstates in $P (N_0 + q_0 , q_0 ) \bigcup_{q'<q_{0}}
 P (N_0 + q, q )$ become:
\bea
&&H_1 \left\{ \vert V_n >  +   \sum_{V_{n'} \in P (N_{0} + q', q'< q_{0})}
 \, {\lambda_{n'}\over \epsilon_n - \epsilon_{n'} }\vert V_{n'} > \right\}
\nonumber\\
&& = \epsilon_n \left\{ \vert V_n > + \sum  {\lambda_{n'}\over \epsilon_n -
\epsilon_{n'}} \vert V_{n'} > \right\}
\eea

A potential problem arises if $\epsilon_{n'} = \epsilon_n$ for some $n'$.
However this would imply for $H_1$ a structure of the form $H_1 \simeq
\epsilon_n \vert V_n><V_n \vert + \epsilon_n ^c \vert V_{n'}><V_{n'} \vert +
\lambda_n \vert V_{n'}><V_n \vert$ which implies that $H_1$ is not
diagonalizable (by Cayley-Hamilton theorem) but since $H_1$ is conjugate to a
hermitian Hamiltonian, this situation cannot occur.

It follows that the computation of all eigenfunctions and eigenvalues of $H_1$
can be done by diagonalizing finite-dimensional matrices and we shall now
present  some specific subsets of eigenfunctions, beginning with the simplest.
We shall only consider here the diagonalization in the space $P(N_0, 0)$.

Polynomials in positive-index variables with fixed $N_V$ and $N_0$ are a finite
 invariant subset of the full Hilbert space.  Two subspaces of $P (N_0 , 0)$
have a particularly simple form for $H_1$.  The first one is the one-matrix,
 zero-vector subspace $N_V = 0$ which was investigated fully in \cite{Je}.
The second one is the subspace of polynomials in $\{ \psi_1^{ab}\}$, which is
easily seen to be invariant.

The pure one-matrix eigenfunctions are the characters of all representations
$R_n $ of $SU(N)$, labeled by a Young tableau $Y_{R_{n}}$.  The corresponding
eigenvalue is given by a simple formula:
\be
\epsilon_Y = {1\over 2} \sum_{\{{\rm lines\,\,} n_{1} \geq n_{2} \geq
\cdots n_{p}\}} \, n_k (n_k - 2k + 1)
\ee
It turns out that it can also be rewritten as:
\be
\epsilon_Y = {1\over 2} n_Y (n_Y - 1) \quad {\chi_Y ( 2 - {\rm transposition})
\over \chi_Y ({\bf 1})}
\ee
where $n_Y$ is the total number of boxes in $Y$; $\chi_Y$ is the character
function of the permutation group over $n$ elements in the representation of
 $S_n$ labeled by $Y$.  In this way a connection is established between
representations of $SU(N)$, permutation groups $S_n$ and eigenfunctions of
 $H_1$.\footnote{We are indebted to Marc Bellon for pointing out this
connection to us.}

The connection is even more apparent in the second sector, namely polynomials
 of degree $n$ in $\psi_1^{ab}$. $H_1$ reduces to:
\be
H_1 = \sum_{b_{1}, b_{2} = 1}^n \quad P_{b_{1} b_{2}} \quad ({\rm permutation
\,\, of\,\, 2\,\, indices \,\, of\,\, b-type)}
\label{30}
\ee
Diagonalizing $H_1$ is now a simple problem in representation theory of
 $S_n$.  First of all, one shows that $H_1$ commutes with $S_n$.

Indeed  $H_1$ commutes with all 2-transpositions.  Consider $P_{a_{1} a_{2}}$.
Contributions to $H_1 \cdot P_{a_{1} a_{2}} - P_{a_{1} a_{2}} \cdot   H_1 $
 only come from the permutations $P_{a_{1} (b_{2} \not= a_{2})}$,  $P_{b_{1}
\not= a_{1} , a_{2}}, P_{a_{1} a_{2}}$ in $H_1$.  $P_{a_{1} a_{2}}$ commutes
 with itself, and it is easy to check that
\be
\left[ P_{a_{1} a_{2}} , P_{a_{1} b} + P_{a_{2} b} \right] = 0
\ee

{}From this and Schur's lemma, it follows that $H_1$ is identical to
 $\lambda_R \cdot {\bf 1}$ on any representation $R$ of $S_n$.  Moreover the
value of $\lambda_R$ is immediately given as $\lambda_R = {{\rm Tr} H_1\over
 {\rm Tr} {\bf 1}}$ and from (\ref{30}) it follows that, given a
representation $R$ with Young tableau $Y_R$, one has
\be
e_Y = {1\over 2} \, n_Y (n_Y - 1 ) \, {\chi_Y (2-{\rm transposition })\over
\chi_Y ({\bf 1})}
\label{32}
\ee
Here $n_Y = n$ is the number of boxes in any Young tableau giving a
 representation of $S_n$.  We recover the same formula as in the pure
one-matrix case, a fact which we shall soon interpret.

The eigenfunctions of (\ref{30}) are constructed in a canonical way as basis
vectors of a given representation $R_Y$ with Young tableau $Y$ \cite{FH}. 
 Given the
lines in $Y$ of length $(n_1 \geq n_2 \geq \cdots n_q)$, one first construct
all ordered $n$-uples, that is, all the ways of inserting the numbers from 1
to $n$  into the boxes of $Y_R$ such that numbers always increase from left to
 right in a line and from top to bottom in a column.  One then defines two
operators:
\bea
P && =  \prod_{{\rm lines}} \, \left\{ {\rm symmetrizing\,\, operator\,\,
over\,\, indices\,\, in\,\, line}\right\}\nonumber\\
Q && = \prod_{{\rm columns}} \left\{ {\rm
antisymmetrizing\,\,operator\,\,over\,\,indices\,\, in\,\, columns }\right\}
\eea
Finally one acts by $P.Q$ (Young symmetrizer), once, on each of the originally ordered $n$-uples.

  Interpretation of the identity of eigenvalues in these two different sectors
 comes from considering the original (unreduced) quadratic Hamiltonian $H_1$.
 $H_1$ is a scalar Hamiltonian, invariant under the action of unitary matrices
 $SU(N)$ on hermitian matrices $M_{ij}$.  From Schur's lemma it follows that
{\it all} its eigenvalues depend only on the choice of a particular
representation
of $SU(N)$ on which it is made to act.  In particular this eigenvalue can be
determined by having $H_1$ act upon an invariant one-dimensional vector space
inside this representation, i.e.  the character of $R_Y$.  Hence the
eigenvalues (\ref{32}) determined in the one-matrix model extend to all
eigenfunctions of $H_1$ identified with explicit matrix elements in a
particular representation.  In particular the polynomials eigenfunctions
of (\ref{8}) are expectation values between particular vectors
$<\bar{x}_{a_{1}} \otimes \cdots\bar{x}_{a_n} \vert$ and
 $\vert x_{b_{n}} \otimes \cdots x_{b_1 >}$ of $M^{\otimes n}$
symmetrized according to a particular $n$-box Young tableau so as
to get an irrep. of $SU(N)$.  Note that indeed $H_1$ does not act
on $x,\bar{x}$ as dynamical variables.

It immediately follows that the study of more complicated eigenstates
combining $\phi_n^0$ and $\psi_n^{ab}$ will generate the same spectrum,
 again with eigenfunctions interpreted as matrix elements - or linear
combinations thereof - of a given representation.  We have been able to
work out all eigenvalues and eigenfunctions up to $N_0 = 4$; the first
examples are presented here:
\bea
N_0 = 1 ,&& N_V = 1:  \psi_1^{ab} , \epsilon = 0\nonumber\\
&&N_V = 0:  \phi_1 , \epsilon = 0\nonumber\\
N_0 = 2 , &&N_V = 0 :  {\rm characters \,\, of\,\, representation\,\,
(\Box )^{\otimes 2}\,\,of} \,\, SU(N);   \epsilon = \pm 2\nonumber\\
&&N_V = 1:  \psi_2^{ab} \pm \psi_1^{ab} \phi_1 ; \quad \epsilon = \pm
2\nonumber\\
&&N_V = 2:  {\rm permutation\,\,representation}\,\,  , \psi_1^{ab} \psi_1^{cd}
\pm \psi_1^{ad} \psi_1^{ab} ; \epsilon = \pm 2 \nonumber\\
N_0 = 3,
&&N_V = 0:  {\rm characters\,\, of\,\, representation\,\, (\Box )^{\otimes 3}
\,\,of \,\,} SU(N) ; \epsilon = \pm 6,0.\nonumber\\
&&N_V = 1 :   \left\{ \psi_3^{ab} - \psi_1^{ab} \phi_2 , \psi_2^{ab} \phi_1^0 -
\psi_1^{ab} \phi_2 \right\} : \epsilon = 0\nonumber\\
&& 2 \psi_3^{ab} \pm 2 \psi_2^{ab} \phi_1 \pm \psi_1^{ab} \phi_2 + \psi_1^{ab}
(\phi_1 )^2: \epsilon = \pm 6\nonumber\\
&&N_V = 2:  \left\{ \psi_2^{ab} \psi_1^{cd} - \psi_1^{ab} \psi_2^{cd} ;
\psi_2^{ab} \psi_1^{cd} - \psi_1^{ab} \psi_1^{cd} \phi_1 \right\} : \epsilon =
0 \nonumber\\
&&\hskip-40pt\left(\psi_2^{ab} \psi_1^{cd} + \psi_1^{cb} \psi_2^{cd}\right)
\pm  \left( \psi_1^{ad} \psi_2^{cb} + \psi_2^{ad} \psi_1^{cb}\right) + \left(
\psi_1^{ad} \psi_1^{cb} \pm \psi_1^{ab} \psi_1^{cd}\right) \phi_1 : \epsilon =
\pm 6 \nonumber\\
&&N_V = 3:  {\rm permutation\,\, representation}\,\,  , \epsilon = \pm 6,0 \,
.\nonumber\\
N_0 = 4,
&&N_V = 0 :  {\rm characters \,\, of}\,\, (\Box )^{\otimes 4}; \quad
\epsilon = \pm 12 , \pm 4,0\nonumber\\
&&N_V = 1:  \epsilon = \,\,0 \,\,{\rm once}, \quad  \pm 4, {\rm twice \,\,
each}, \pm 12 \,\, {\rm once}\nonumber\\
&&N_V = 2:  {\rm two\,\, types\,\, of\,\, eigenstates}:\nonumber\\
&& ({\rm with \,\,antisymmetrized\,\,vector\,\,index }):
 \epsilon = 0 , \pm 4 \;({\rm twice}) , - 12 \nonumber\\
&& ({\rm with \,\,symmetrized\,\, vector\,\, index}): \epsilon = 0, \pm 4 ;
({\rm twice}), + 12  \nonumber
\eea

The eigenvalues are defined up to a shift by $2N_0$ $\phi_0$, since $\phi_0$
is in fact not a variable but a $c$-number which may be given formally any
 value.  We have not given the forms of eigenfunctions for $N_0 = 4$ since
they are quite cumbersome and, due to the degeneracy, not canonical anyway.

We may now apply the diagonalizing scheme developed above to obtain
eigenfunctions containing both positive and negative indices, but we shall
leave this rather technical problem for the moment. An interesting extension
of our discussion is now the introduction of interaction terms in the
matrix-vector model, leading to external potential terms in the collective
hamiltonian. In particular the construction of exactly algebraically solvable
theories should be possible on similar schemes to the one in \cite {AJ2}.
The relevant algebra is here the non-linear structure constructed
in \cite {AB}.

\medskip
\section{ The Classical Collective Field Theory}
\medskip
We now discuss a collective description of the classical spin Caloger Moser
model, in order to illuminate the basic properties of the theory. 
We consider here the real vector case.
After fixing the momentum map the reduced matrix model
is written in terms of the eigenvalues $\lambda_i, i=1\cdots N$ and the
vector degrees of freedom as
\be
H_1 = - \sum_{i=1}^N \, {\partial^2\over \partial\lambda_i^2} + \sum_{i\not=j}
\, {Q_{ij} Q_{ji}\over (\lambda_i - \lambda_j )^2}
\ee
where the charges
\be
Q_{ij} = \sum_{b=1}^{r} a_i^b a^{b \dagger }_j
\ee
close a $U(N)$ algebra.  This is rewritten as
\be
H_1 = - {1\over 2} \sum_{i=1}^N \, {\partial^2\over \partial\lambda_i^2} +
\sum_{i<j} \, {q_i^{ab} q_j^{ba}\over (\lambda_i - \lambda_j)^2}
\ee
with the generators
\be
q_i^{bc} = a_i^b a^{c \dagger }_i
\ee
realizing a $U(r)$ algebra at each site $i$.  The classical transition to
collective variables is now done with
\bea
\phi(x) & = & \sum_{i=1}^N \, \delta (x-\lambda_i )\\
\phi \partial \pi & = & \sum_i \, p_i \, \delta (x-\lambda_i )
\eea
describing the scalar density  (and current) and
\be
{\cal J}^{ab} (x) = \sum_{i=1}^N \, q_i^{ab} \, \delta (x-\lambda_i )
\ee
representing a local $U(r)$ current algebra. A simple rewritting of the
Hamiltonian leads us to
\be
H_i = \int dx {1\over 6} \left( \alpha_+ (x)^3 - \alpha_- (x)^3 \right) +
{1\over 2} \int dx \int dy \, {{\cal J } ^{ab} (x) {\cal J}^{ba} (y)\over
(x - y)^2}
\ee
with $\alpha_{\pm} = \phi \pm \partial \pi$.  The first term is the
standard bosonic Hamiltonian for the eigenvalue density \cite{Je} while
the second term gives the $U(r)$ current algebra Hamiltonian of the form
\be
\sum_k \, k \, {\cal J}^{ab} (k) \, {\cal J}^{ba} (-k)
\ee
coinciding with the current-current interaction in \cite{HT}.
These two terms are apparently decoupled. However one can easily check
the Poisson bracket structure:
\bea
\left\{ \phi \partial \pi (x)  ,  {\cal J}^{ab} (y) \right\} & = &
\partial_x \left( {\cal J}^{ab} (y) \delta (x-y) \right)\\
\left\{ \phi (x)  ,   {\cal J}^{ab} (y) \right\} & = & 0
\label{pb}
\eea
Consequently the interaction between the momentum density $\pi (x)$ and
the current ${\cal J}^{ab} (x)$ arises through the Poisson structure.
This situation brings to mind the problem which arises when seeking to add a
spin-type variable to the relativistic (Ruijsenaars-Schneider \cite{RS})
Calogero-Moser system:
such an addition may not be achieved while keeping the spin variable
independent (in the Poisson structure) of the position/momenta variables.
The relation might be deeper actually since the commuting quantum spin
Hamiltonians constructed from the quantum $R$ matrix structure of the
Calogero-Moser system are in fact
closer to a Ruijsenaars-Schneider-like system.

A canonical transformation of the fields so as to undo the ``off-diagonal''
coupling in (\ref{pb}) will then induce an explicit coupling term in the
hamiltonian, of the same form as in (\ref{ham2}).

The currents themselves obey the algebra
\be
\left\{ {\cal J}^{ab} (x) , {\cal J}^{cd} (y) \right\} = \left( {\cal J}^{ad}
 (x) \delta_{bc} - \delta_{ad} {\cal J}^{cb} (x) \right) \delta (x-y )
\ee
without a central charge.

It is clear that in quantization a central charge will arise.  Considering
 the filling of the ground state near the upper and lower Fermi momenta,
 we have
\be
\left\{ \alpha_{\pm} (x) , \alpha_{\pm} (y) \right\} = \pm 2 \, \partial
 \delta (x-y)
\ee
while for $\phi (x) = {1\over 2} (\alpha_+ - \alpha_- )$ the anomaly cancels.
 Similarly we then expect that
\be
{\cal J}^{ab} (x) = J^{ab} (x) - \tilde{J}^{ab} (x)
\ee
where $J^{ab} (x)$ and $\tilde{J}^{ab} (x)$ obey $U(r)$ Kac-Moody algebras with
opposite central charges $k$.

The complex vector case may also be treated in this way. Indeed the generators
$F_{ij}$  in (1) also admit an oscillator-type representation corresponding to
internal symmetry indices of a vector representation of $U(r) \times U(r)$.
Precisely one
defines the variables:

\be A_{i}^{a;1,2} \equiv (x_i^a , -\bar{p}_i^a) \; ; \;
    A_{i}^{\dagger \, a;1,2} \equiv (p_i^a , -\bar{x}_i^a)
\nonumber
\ee

and the algebraic structure of the classical theory is the same as in the
real vector case, although with different reality properties for the currents.

Let us finally comment on another approach, devised independently 
in \cite{Awa2}. 
This work uses a different set of collective variables taken to be the 
spin version of standard densities. In our approach the $SU(r)$ degrees of
freedom are dealt with by current algebra fields, as in \cite{HT,Sc}.
A possible comparison therefore could be made from a detailed
study of the spectrum in both representations.
 
\vskip0.5in
{\bf Acknowledgements}

This work was sponsored by CNRS and CNRS-NSF Exchange Programme AI 06-93 (J.A.);
and DOE Grant DE-FG02-91ER-40688, Task A, Brown University. J.A. wishes to
thank Brown University Physics Department members for their kind support.

\end{document}